\documentstyle[aps]{revtex}

\draft

\begin{document}

\title{Quantum distributions prescribed by factorization hypothesis of probability}

\author{Qiuping A. Wang}
\address{Institut Sup\'erieur des Mat\'eriaux du Mans,\\ 44, Avenue F.A.
Bartholdi, 72000 Le Mans, France}

\date{\today}

\maketitle

\begin{abstract}
Nonextensive quantum gas distributions are investigated on the
basis of the factorization hypothesis of compound probability
required by thermodynamic equilibrium. It is shown that the
formalisms of Tsallis nonextensive statistical mechanics with
normalized average give distribution functions for standard
bosons and fermions obeying Pauli principle. The formalism with
unnormalized average leads to a intermediate quantum distribution
comparable to that of fractional exclusion statistics, with Fermi
surface at $T=0$ depending on the parameter $q$.
\end{abstract}

\pacs{05.20.-y;05.30.-d;05.30.Pr}

\section{introduction}
It is believed that long-range correlations can often (if not
always) be related to fractal or chaotic space-time structures and
also to power laws and nonextensive
phenomena\cite{Krog00,Nasc00,Nott66,Ahme00,Mand97,Tsal95}. That
is why the fractal inspired nonextensive statistical mechanics
($NSM$)\cite{Tsal88} is receiving much attention and widely
considered as a possible valid theory to describe systems with
complex correlations. Considering the fractal or chaotic
behaviors and nonextensive or fractional effects in quantum
systems like, for example, correlated electrons in
superconductor, solitons and quasi-particles in condensed matters
\cite{Krog00,Badi99,Wu94,Hald91,Mont93}, we naturally expect a
quantum statistical mechanics within $NSM$. One of the first steps
in this direction is taken by B\"uy\"ukkili\c{c} et
al\cite{Buyu93} in giving the generalized quantum distributions
with the so called factorization approximation.

The derivation of the conventional quantum one-body distribution
from the compound many-body distribution is straightforward within
Boltzmann-Gibbs statistics because the latter considers only short
range interactions with additive entropy and energy. This allows
the many-body density operator $\rho$ to be factorized into the
product of one-body density operators $\rho_i$ of $N$ particles
of the system :
\begin{eqnarray}                                        \label{1}
\rho=\prod_{i=1}^N\rho_i.
\end{eqnarray}
But in nonextensive statistical mechanics, this kind of
factorization of compound probability is impossible if we still
consider, for independent particles or ideal gas, extensive energy
with total hamiltonian $H$ given by
\begin{equation}                                        \label{2}
H=\sum_{i=1}^N H_i=\sum_{i=1}^N(\frac{p_i^2}{2m}+V_i),
\end{equation}
where $H_i=\frac{p_i^2}{2m}+V_i$ is the one-body hamiltonian of
$i^{th}$ particle, $p_i$ its momentum and $V_i$ its potential
energy. To overcome this difficulty, a so called factorization
approximation is proposed\cite{Buyu93} which imposes Eq.(\ref{1})
in keeping Eq.(\ref{2}). This approximation allows to obtain a
nonextensive quantum distribution of $NSM$ with complete
distribution (i.e. $\texttt{Tr}\rho=1$) and unnormalized average
(i.e. $\bar{x}=\texttt{Tr}\rho^q x$ where $x$ is certain operator
\cite{Tsal88}). It is, explicitly or not, employed in most of the
applications of $NSM$ to the cases like, among others, boson and
fermion systems\cite{Buyu93,Buyu00,Kani96}, the polytropic model
of galaxies, solar neutrinos, peculiar velocity of galaxy
clusters, electron plasma turbulence (for updated comments on
these works, we refer to reference \cite{Tsal01}), and the
application of nonextensive blackbody distribution to laser
physics\cite{Wang98}. Although these applications clearly
evidence the existence and the important role of the
$q$-exponential distribution in the nature, the justification of
the passage from many body $q$-distribution to one-body one is
missing. Recently, a direct analysis\cite{Wang01} shows that,
assuming Eq.(\ref{1}) and Eq.(\ref{2}) at the same time means
neglecting a correlation energy which is in general not
negligible. A detailed computation for $N$-oscillator
system\cite{Lenz01} also shows that the partition function given
on the basis of Eq.(\ref{1}) is completely different from that
given by using Eq.(\ref{2}).

In this paper, the factorization hypothesis is discussed from a
new point of view relative to Abe's general
pseudoadditivity\cite{Abe01} for nonextensive systems in thermal
equilibrium. This discussion leads to the conclusion that
Eq.(\ref{1}) must be respected in all coherent and exact
treatments of equilibrium systems. On this basis, quantum
distributions are derived within different formalisms of $NSM$ :
1) complete distribution ($\texttt{Tr}\rho=1$) with the
normalized average $\bar{x}=\texttt{Tr}\rho x$\cite{Tsal88}; 2)
incomplete distribution with $q$-normalization
$\texttt{Tr}\rho^q=1$ and the normalized average
$\bar{x}=\texttt{Tr}\rho^q x$\cite{Wang01}; 3) complete
distribution with unnormalized average
$\bar{x}=\texttt{Tr}\rho^qx$\cite{Tsal88}.

\section{Thermal equilibrium and factorization of compound
probability}

In Boltzmann-Gibbs statistics, Eq.(\ref{1}) is a natural result
of the short-range interactions or of the independence of the
subsystems which necessarily lead to extensive entropy $S$ and
internal energy $E$. With the temperature definition
$1/T=\beta=\frac{\partial S}{\partial E}$ (Boltzmann constant
k=1), the extensivity of $S$ and $E$ allows to verify the zeroth
law or the existence of thermal equilibrium. So, the uniqueness
of Gibbs-Shannon entropy $S=-\sum_ip_i\ln p_i$ subject to
Shannon-Khinchin axioms\cite{Sant97} can be regarded from a
different viewpoint. We can also state that, given Gibbs-Shannon
entropy, the existence of thermal equilibrium necessarily leads
to the additivity of entropy or to Eq.(\ref{1}). This statement
seems self-evident and useless for Boltzmann-Gibbs statistical
mechanics, because Eq.(\ref{1}) and the axiom about the
additivity of entropy are obvious in view of the independence of
subsystems. But it is not the case in $NSM$. The pseudoadditivity
axiom $S=S_1+S_2+(1-q)S_1S_2$\cite{Tsal88} for a composite system
containing two subsystems $1$ and $2$ with respectively entropies
$S_1$ and $S_2$, is in fact a consequence of Eq.(\ref{1})
supposed Tsallis entropy $S=-\frac{1-\texttt{Tr}\rho^q}{1-q}$. As
a consequence, the nonextensive theory is entirely subject to the
``independence" of subsystems. The problem is that, in
nonextensive physics, the subsystems are in general dependent on
each other. So Eq.(\ref{1}) or the pseudoadditivity is in this
case somewhat artificial and seems to hide something which may
restrict the validity of the theory. In many applications of
$NSM$, Eq.(\ref{1}) is sometimes respected, sometimes not,
sometimes combined with the incompatible extensive hamiltonian
Eq.(\ref{2})\cite{Buyu93,Lenz01,Naka02}. This confused situation
is in our opinion due to the fact that Eq.(\ref{1}) is not clearly
founded for nonextensive statistics subject to Tsallis entropy.

Very recently, Abe\cite{Abe01} proposed a general
pseudoadditivity for entropy required by the existence of thermal
equilibrium in composite nonextensive systems :
$f(S)=f(S_1)+f(S_2)+\lambda f(S_1)f(S_2)$ where $f$ is certain
differentiable function satisfying $f(0)=0$ and $\lambda$ a
constant depending on the nature of the system of interest. So
for a system containing $N$ subsystems, the thermal equilibrium
requires following additivity :
\begin{eqnarray}                                        \label{2b}
\ln[1+\lambda f(S)]=\sum_{i=1}^N\ln[1+\lambda f(S_i)].
\end{eqnarray}
On the other hand, Eq.(\ref{2b}) applied to Tsallis entropy means
$f(S)=S$ and $\lambda=1-q$\cite{Abe01}, which directly leads to
$\ln\texttt{Tr}\rho^q=\sum_{i=1}^N\ln\texttt{Tr}\rho_i^q$ or
Eq.(\ref{1}). This interesting result raises Eq.(\ref{1}) from
the level of a special assumption for ``statistically
independent" subsystems to the level a general theorem of
equilibrium thermodynamics with either Boltzmann-Gibbs or Tsallis
entropy. We does not need any more the independence of subsystems
to write Eq.(\ref{1}) which must be respected by all probabilities
of the systems (extensive or nonextensive) that can have thermal
equilibrium. Consequently, all calculations based on the
factorization theorem or using one-body $q$-exponential
distribution are, as a matter of fact, correct and exact
applications of $NSM$. And all calculations of $NSM$ based on
Eq.(\ref{2}) [not compatible with Eq.(\ref{1})] should now be
regarded as ``extensive energy approximation" which may be very
different from the exact treatment and should be employed with
great care.

In what follows, we discuss the quantum distributions proposed by
B\"uy\"ukkili\c{c} {\it et al}\cite{Buyu93} without any
approximation. On the basis of the factorization theorem
discussed above, the derivation of this kind of distribution is
straightforward.

\section{Complete distribution formalism}
Let us begin with the complete distribution ($\texttt{Tr}\rho=1$)
and the normalized average $\bar{x}=\texttt{Tr}\rho x$. This
formalism of $NSM$ is first proposed by Tsallis\cite{Tsal88} and
has received little attention due to some problems\cite{Tsal88}
(e.g. missing Legendre transformation and zeroth law,
thermodynamic stability problem discussed by assuming extensive
energy\cite{Rams95}). Since several years, $NSM$ has much
evolved. Legendre transformation and the zeroth law in this
formalism can be established by using the method of the
references\cite{Wang01}. The stability problem must be revisited
with nonextensive energy satisfying Eq.(\ref{1}). So this
standard formalism of probability theory remains a possible
choices of $NSM$ and still deserves to be studied. The reader
will see below that the formalisms with normalized average can
give the very distribution functions of B\"uy\"ukkili\c{c} {\it
et al}.

For nonextensive quantum gas, we have\cite{Tsal88,Buyu93,Curi96} :
\begin{eqnarray}                                        \label{3}
\rho=\frac{1}{Z}[1-(q-1)\beta(H-\mu N)]^\frac{1}{q-1}
=\frac{1}{Z}e^{-\beta h'}
\end{eqnarray}
where $h'=\frac{\ln[1+(1-q)\beta(H-\mu N)]}{\beta(1-q)}$ can be
called ``deformed hamiltonian". The grand partition function $Z$
is then given by
\begin{eqnarray}                                        \label{4}
Z &=&\texttt{Tr}[1-(q-1)\beta(H-\mu N)]^\frac{1}{q-1}
\\\nonumber &=&\texttt{Tr}e^{-\beta h'}.
\end{eqnarray}
From the factorization theorem, we can write
\begin{eqnarray}                                        \label{4a}
Z &=&\texttt{Tr}e^{-\beta \sum_nh'_i} \\\nonumber
&=&\prod_k\sum_{n_k}e^{-n_k\beta\epsilon_k}
\\\nonumber &=&\prod_k\sum_{n_k}e^{-\beta n_k\epsilon_k}
\end{eqnarray}
where $\epsilon_k$ is the eigenvalue of
$h'_i=\frac{\ln[1+(1-q)\beta(H_i-\mu)]}{\beta(1-q)}$, the deformed
one-particle hamiltonian satisfying $h'=\sum_{i=1}^N h'_i$, and
$n_k$ the occupation number of the one-particle state $k$. For
boson and fermion, we obtain, respectively,
\begin{eqnarray}                                        \label{5}
Z=\prod_k\sum_{n_k=0}^\infty e^{-n_k\beta\epsilon_k}
=\prod_k\frac{1}{1-e^{-\beta\epsilon_k}} \;\;\;and\;\;\;
Z=\prod_k\sum_{n_k=0}^1 e^{-n_k\beta\epsilon_k}
=\prod_k(1+e^{-\beta\epsilon_k}).
\end{eqnarray}
Then, it is straightforward to show that, just as in the
conventional quantum statistics ,
\begin{eqnarray}                                        \label{6}
\bar{n}_l=\texttt{Tr}\rho n_l=-\frac{1}{\beta}\frac{\partial (\ln
Z)}{\partial\epsilon_l}=\frac{1}{e^{\beta\epsilon_l}\pm1}
=\frac{1}{[1+(1-q)\beta(e_l-\mu)]^{\frac{1}{1-q}}\pm1}
\end{eqnarray}
where $e_l$ is the eigenvalue of the one-particle hamiltonian
$H_i$. ``+" and ``-" correspond to fermion and boson,
respectively. This result is just that given by B\"uy\"ukkili\c{c}
{\it et al}\cite{Buyu93} with only a change
$(q-1)\rightarrow(1-q)$.

\section{Incomplete distribution formalism}
For incomplete distribution with the $q$-normalization, the Grand
partition function $Z$ is given by
\begin{eqnarray}                                        \label{7}
Z^q &=&\texttt{Tr}[1-(1-q)\beta(H-\mu N)]^\frac{q}{1-q}
\\\nonumber &=&\texttt{Tr}e^{-q\beta h'} \\\nonumber
&=&\texttt{Tr}e^{-q\beta \sum_nh'_i} \\\nonumber
&=&\prod_k\sum_{n_k}e^{-qn_k\beta\epsilon_k}
\\\nonumber &=&\prod_k\sum_{n_k}e^{-q\beta n_k\epsilon_k}.
\end{eqnarray}
The deformed hamiltonians $h'$ and $h'_i$ are the same functions
as given above with the transform $(q-1)\rightarrow
(1-q)$\cite{Wang01}. The same machinery leads to
\begin{eqnarray}                                        \label{8}
\bar{n}_l=\texttt{Tr}\rho^qn_l=\frac{1}{e^{q\beta\epsilon_l}\pm1}
=\frac{1}{[1+(q-1)\beta(e_l-\mu)]^{\frac{q}{q-1}}\pm1}.
\end{eqnarray}
These distributions are equivalent to Eq.(\ref{6}) with only a
difference in the $q$-dependence. Eq.(\ref{6}) and Eq.(\ref{8})
represent standard bosons and fermions satisfying Pauli exclusion
principle. For example, $n_l\leq 1$ for fermions. It should be
noticed that, in this case, the Fermi surface at $T=0$ or
$\beta=\infty$ is the same as in the conventional case and
independent of the parameter $q$.

\section{Complete distribution with unnormalized average}

On the other hand, in the formalism of complete distribution with
unnormalized average, the things will be different because this
average, unlike the normalized ones, can lead to nonadditive
deformed internal energy even with additive deformed hamiltonian
$h'$. This formalism received a lot of attention and finally was
found to show some puzzling properties\cite{Tsal88}. But up to
now, there is no really solid reason for abandoning it definitely.
We still see various kind of unnormalized expectations widely
used in many fields, especially in financial
problems\cite{Tsal95,Ante01}. It is also found that this
formalism is the only one of $NSM$ that can give standard
statistical interpretation of heat and work\cite{Wang01b}. So it
is still of interest to study this formalism in the circumstance
of quantum systems.

The partition function is still given by Eq.(\ref{4}) with a
transform $(q-1)\rightarrow (1-q)$\cite{Tsal88}. The occupation
number is calculated as follows
\begin{eqnarray}                                        \label{9}
\bar{n}_l &=&\texttt{Tr}\rho^q n_l  \\\nonumber &=&\texttt{Tr}
n_l\frac{1}{Z^q}[1+(q-1)\beta(H-\mu N)]^{\frac{q}{1-q}}
\\\nonumber
&=&\frac{1}{Z^q}\prod_k\sum_{n_k} n_le^{-qn_k\beta\epsilon_k}
\\\nonumber &=&\frac{-1}{q\beta Z^q}\frac{\partial}{\partial
\epsilon_l}\prod_k\sum_{n_k} e^{-qn_k\beta\epsilon_k}
\\\nonumber &=&-\frac{1}{Z^q}\prod_k\sum_{n_k}
e^{-qn_k\beta\epsilon_k}\frac{1}{e^{q\beta\epsilon_l}-1}
\\\nonumber
&=&\texttt{Tr}\rho^q\frac{1}{e^{q\beta\epsilon_l}-1}
\\\nonumber
&=&\frac{1}{e^{q\beta\epsilon_l}-1}
\prod_k\frac{(1-e^{-\beta\epsilon_k})^q}{1-e^{-q\beta\epsilon_k}}
\end{eqnarray}
for bosons and
\begin{eqnarray}                                        \label{10}
\bar{n}_l
&=&\frac{1}{e^{q\beta\epsilon_l}+1}\prod_k\frac{1+e^{-q\beta\epsilon_k}}
{(1+e^{-\beta\epsilon_k})^q}
\end{eqnarray}
for fermions. Eq.(\ref{9}) and (\ref{10}) can be written as
\begin{eqnarray}                                        \label{11}
\bar{n}_l =\frac{Q_{\pm}}{e^{q\beta\epsilon_l}\pm 1}
=\frac{Q_{\pm}}{[1+(q-1)\beta(e_l-\mu)]^{\frac{q}{q-1}}\pm1}
\end{eqnarray}
where $Q_+=\texttt{Tr}\rho^q=\prod_k\frac{1+e^{-q\beta\epsilon_k}}
{(1+e^{-\beta\epsilon_k})^q}$ (or
$Q_-=\prod_k\frac{(1-e^{-\beta\epsilon_k})^q}
{1-e^{-q\beta\epsilon_k}}$) can be regarded as a parameter
depending on $q$. $Q_{\pm}>1$ $Q_{\pm}=1$ and $Q_{\pm}<1$ for
$q<1$, $q=1$ and $q>1$, respectively. These distributions seems
interesting because they allow intermediate occupation number
between that of bosons and fermions. In particular, at absolute
zero, for ``fermionlike" particle with ``+" in Eq.(\ref{11}),
$\bar{n}_l=Q$ when $e_l<\mu$ and $\bar{n}_l=0$ when $e_l>\mu$.
This means that it is possible for several ``fermions" to occupy
an one-particle quantum state if $Q>1$ or $q<1$. Consequently,
the Fermi surface $\epsilon_F$ at $T=0$ or $\beta=\infty$ changes
as a function of $Q$ or of the interaction between the particles
: $\epsilon_F=\frac{\varepsilon_F}{Q^{2/3}}$ where $\varepsilon_F$
is the conventional Fermi energy at $q=1$ or $Q=1$. This result
can be compared to that of the {\it fractional exclusion
statistics}\cite{Wu94} for intermediate particles different from
bosons and fermions. It is not surprising to see that
nonextensive statistics has similar effect to that of fractional
statistics describing interacting particle or elementary
excitation\cite{Wu94,Hald91}, because the $q$-distribution is
nothing but a result of long range interactions. However, the
fact that only the $NSM$ formalism with unnormalized average can
give the intermediate quantum distributions seems to deserve
further investigation. Quite interesting efforts have been made
by some authors\cite{Kani96,Buyu01} to relate nonextensive
statistics and the quantum distributions given by Eq.(\ref{6}) or
(\ref{8}) to fractional exclusion statistics.

It is obvious that if we assume the $q$-normalization for
incomplete distribution, $Q=1$ and Eq.(\ref{11}) will become
Eq.(\ref{8}).

\section{Conclusion}
Summing up, on the basis of the factorization theorem of compound
probability prescribed by thermodynamic equilibrium, the quantum
gas distributions within different formalisms of $NSM$ are
investigated. It is shown that only the formalisms with normalized
average can give distribution functions for standard bosons and
fermions. The formalism with unnormalized average leads to a
intermediate distribution similar to that of fractional exclusion
statistics, with Fermi surface at $T=0$ depending on the
parameter $q$. This different quantum properties of $NSM$ due to
normalizations remains something to be understood. A detailed
study of the relation between the unnormalized quantum
distribution and the fractional exclusion statistics would be of
interest.


\begin{thebibliography}{99}
\bibitem {Krog00}
Kr\"oger H, Fractal geometry in quantum mechanics field theory
and spin systems, {\em Physics Report,\/} 2000;323(2):82-181
\bibitem {Nasc00}
El Naschie MS, On the elctroweak mising angle in $E^\infty$ {\em
Chaos, Solitons $\&$ Fractals,\/} 2000;11(11):1803-1807
\bibitem {Nott66}
Nottale L, Scale relativity and fractal space-time {\em Chaos,
Solitons $\&$ Fractals,\/} 1966;7(6):877-938
\bibitem {Ahme00}
Ahmed E, Hegazi AS, On infinitesimally deformed algebra and
fractal space-time theory {\em Chaos, Solitons $\&$ Fractals,\/}
2000;11(11):1759
\bibitem {Mand97}
Mandelbrot B., Fractals in Finance, Springer, Berlin, 1997
\bibitem {Tsal95}
C. Tsallis, {\em Chaos, Solitons $\&$ Fractals,\/} 1995;6:539.
\bibitem {Tsal88}
Tsallis C, {\em J. Statis. Phys.,\/} 1988;52:479; Curado E.M.F.,
and Tsallis C., {\em J. Phys. A: Math. Gen.\/} 1991;24:L69;
Tsallis C, Mendes R.S. and Plastino A.R., {\em Physica A,\/}
1998;261:534
\bibitem {Badi99}
Badiali J.P., {\em Phys. Rev. E,\/} 1999;60:2533
\bibitem {Wu94}
Wu Y.S., {\em Phys. Rev. Lett.,\/} 1994;73:922
\bibitem {Hald91}
Haldane F.D.M., {\em Phys. Rev. Lett.,\/} 1991;67:937
\bibitem {Mont93}
Monteiro MR, Roditi I., {\em Mod. Phys. Lett. B,\/} 1993;7:1843
\bibitem {Buyu93}
B\"uy\"ukkili\c{c} F. and Demirhan D., {\em Phys. Lett. A,\/}
1993;181:24; B\"uy\"ukkili\c{c} F., Demirhan D. and G\"ule\c{c}
A., {\em Phys. Lett. A,\/} 1995;197:209; Tirnakli U.,
B\"uy\"ukkili\c{c} F. and Demirhan D., {\em Physica A,\/}
1997;240:657; Tirnakli U., B\"uy\"ukkili\c{c} F. and Demirhan D.,
{\em Phys. Lett. A ,\/} 1998;245:62; Torres D.F. and Tirnakli U.,
{\em Physica A,\/} 1998;261:499; Tirnakli U. and Torres D.F., {\em
Physica A,\/} 1999;268:152
\bibitem {Buyu00}
B\"uy\"ukkili\c{c} F. and Demirhan D., {\em Euro. Phys. J. B,\/}
2000;14:705
\bibitem {Kani96}
Kaniadakis G., Lavagno A. and Quarati P., {\em Modern Phys. Lett.
B,\/} 1996;10:497; Kaniadakis G., Lavagno A. and Quarati P., {\em
Nucl. Phys. A,\/} 2001;680:94c-97c
\bibitem {Tsal01}
Tsallis C., {\em Entropic nonextensivity : a possible measure of
complexity,}{\em Chaos, Solitons $\&$ Fractals,\/} 2001;13:371 and
e-print, cond-mat/0010150; Tsallis C., {\em Nonextensive
statistics : theoretical, experimental and computational
evidences and connections,}{\em Brazilian J. Physics,\/} 1999;29:1
\bibitem {Wang98}
Wang Q.A. and Le M\'ehaut\'e A., {\em Phys. Lett. A.,\/}
1998;242:301
\bibitem {Wang01}
Wang Q.A., Pezeril M., Nivanen L., Le M\'ehaut\'e A., {\em Chaos,
Solitons $\&$ Fractals,\/} 2002;13:131, e-print :
cond-mat/0010294; Wang Q.A., {\em Chaos, Solitons $\&$
Fractals,\/} 2001;12:1431, e-print, cond-mat/0009343
\bibitem {Lenz01}
Lenzi E.K., Mendes R.S., da Silva L.R. and Malacarne L.C., {\em
Physics Lett. A.,\/} 2001;289:44
\bibitem {Sant97}
dos Santos R.J.V, {\em J. Math. Phys.,\/} 1997;38:4104; Abe S.,
{\em Physics Lett. A.,\/} 2000;271:74
\bibitem {Naka02}
Nakamichi A., Joichi I., Iguchi O. and Morikava M., {\em Chaos,
Solitons $\&$ Fractals,\/} 2002;13:595
\bibitem {Abe01}
S. Abe, {\em Phys. Rev.E,\/}{\bf 63}(2001)061105; Qiuping A.
Wang, Laurent Nivanen, Alain Le M\'ehaut\'e and Michel Pezeril,
{\em Note on Abe's general pseudoadditivity for nonextensive
systems,} e-print : cond-mat/0111541 (submitted)
\bibitem {Curi96}
Curilef S., Z. Phys. B, 1996;100:433
\bibitem {Rams95}
Ramshaw John D., {\em Phys. Lett. A.,\/} 1995;198:119; Guerberoff
G.R. and Raggio G.A., {\em Phys. Lett. A.,\/} 1998;214:313
\bibitem {Ante01}
C. Anteneodo, C. Tsallis and A.S. Martinez, e-print :
cond-mat/0109203
\bibitem {Wang01b}
Wang Q.A., {\em Unnormalized nonextensive expectation value and
zeroth law of thermodynamics,} e-print : cond-mat/0111238
(submitted)
\bibitem {Buyu01}
B\"uy\"ukkili\c{c} F., H. Uncu and D. Demirhan, {\em
Thermodynamics of the g-on gas,\/} eprint: cond-mat/0103560

\end{thebibliography}
\end{document}